\documentclass{SciPost}

\hypersetup{
    colorlinks,
    linkcolor={red!50!black},
    citecolor={blue!50!black},
    urlcolor={blue!80!black}
}

\usepackage[bitstream-charter]{mathdesign}
\DeclareSymbolFont{usualmathcal}{OMS}{cmsy}{m}{n}
\DeclareSymbolFontAlphabet{\mathcal}{usualmathcal}

\fancypagestyle{SPstyle}{
\fancyhf{}
\lhead{\colorbox{scipostblue}{\bf \color{white} ~SciPost Physics Proceedings }}
\rhead{{\bf \color{scipostdeepblue} ~Submission }}

\fancyfoot[C]{\textbf{\thepage}}
}

\begin{document}

\pagestyle{SPstyle}

\begin{center}{\Large \textbf{\color{scipostdeepblue}{
Differential cross section measurements of top quark pair production for variables of the dineutrino system with the CMS experiment
}}}\end{center}

\begin{center}\textbf{
Sandra Consuegra Rodríguez\textsuperscript{1$\star$}}
\\
\vspace{0.3cm}
on behalf of the CMS Collaboration
\end{center}

\begin{center}
{\bf 1} RWTH Aachen University, Aachen, Germany
\\[\baselineskip]
$\star$ \href{mailto:sandra.consuegra.rodriguez@rwth-aachen.de}{\small sandra.consuegra.rodriguez@rwth-aachen.de}\,\quad
\end{center}

\definecolor{palegray}{gray}{0.95}
\begin{center}
\colorbox{palegray}{
  \begin{tabular}{rr}
  \begin{minipage}{0.36\textwidth}
    \includegraphics[width=60mm,height=1.5cm]{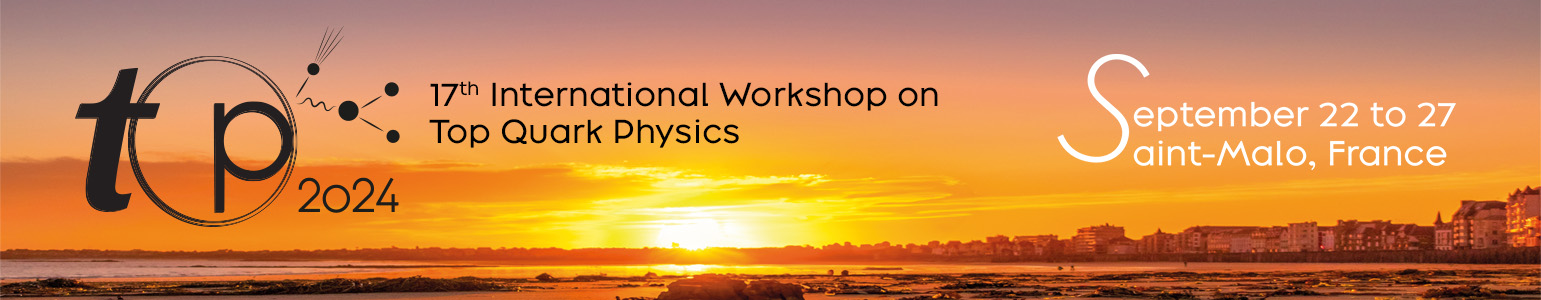}
  \end{minipage}
  &
  \begin{minipage}{0.55\textwidth}
    \begin{center} \hspace{5pt}
    {\it The 17th International Workshop on\\ Top Quark Physics (TOP2024)} \\
    {\it Saint-Malo, France, 22-27 September 2024
    }
    \doi{10.21468/SciPostPhysProc.?}\\
    \end{center}
  \end{minipage}
\end{tabular}
}
\end{center}

\section*{\color{scipostdeepblue}{Abstract}}
\textbf{\boldmath{%
Differential top quark pair cross sections are measured in the dilepton final state as a function of kinematic variables associated to the dineutrino system. The measurements are performed making use of the Run 2 dataset collected by the CMS experiment at the CERN LHC collider, corresponding to proton-proton collisions recorded at center of mass energy of 13 TeV and an integrated luminosity of 138 fb$^{-1}$. The measured cross sections are found in agreement with theory predictions and Monte Carlo simulations of standard model processes.}}

\vspace{\baselineskip}

\noindent\textcolor{white!90!black}{%
\fbox{\parbox{0.975\linewidth}{%
\textcolor{white!40!black}{\begin{tabular}{lr}%
  \begin{minipage}{0.6\textwidth}%
    {\small Copyright attribution to authors. \newline
    This work is a submission to SciPost Phys. Proc. \newline
    License information to appear upon publication. \newline
    Publication information to appear upon publication.}
  \end{minipage} & \begin{minipage}{0.4\textwidth}
    {\small Received Date \newline Accepted Date \newline Published Date}%
  \end{minipage}
\end{tabular}}
}}
}

\vspace{10pt}
\noindent\rule{\textwidth}{1pt}
\tableofcontents
\noindent\rule{\textwidth}{1pt}

\section{Introduction}
\label{sec:introduction}
Precision measurements of top quark pair production provide stringent tests for the validity of the standard model (SM) and play a crucial role in the search for new phenomena. The data sample collected at the CERN LHC has allowed several measurements of differential cross sections for various decay channels and different center-of-mass energies, with measurements often performed as a function of kinematic observables of the visible part of the event (e.g. jets or charged leptons) and intermediate particles [e.g. (anti-)top quark or W boson] \cite{ATLAS:2020ccu,CMS:2021vhb}. In the context of some beyond the standard model (BSM) scenarios the invisible part of the event is modified, therefore, precise and direct measurements of undetected particles in the event (e.g. neutrinos) become quite relevant in the search for new phenomena. The production of a hypothetical top squark pair, in which both top squarks decay to a top quark and a neutralino constitutes an example of such BSM scenarios \cite{CMS:2020pyk}. \\
The physics analysis portrayed in this contribution,  performed making use of proton-proton collision data recorded by the CMS detector \cite{CMS:2008xjf,CMS:2023gfb}, presents the first measurement of top quark pair production differential cross sections as a function of the transverse momentum of the dineutrino system \ensuremath{p_{\text{T}}^{\nu\nu}}, the minimum azimuthal distance between \ensuremath{\vec{p}_{\text{T}}^{\nu\nu}} and leptons \ensuremath{\text{min}[\Delta\phi(\vec{p}_{\text{T}}^{\nu\nu},\vec{p}_{\text{T}}^{\ell})]}, and the two-dimensional (2D) measurement of both observables \cite{CMS:2024tgh}. The selection of these observables is driven by a distinction between SM processes and potential BSM scenarios with comparable signature but including additional sources of undetected particles. The focus on observables related to the dineutrino system and the use of a dedicated deep neural network (DNN) regression to improve the transverse momentum resolution for dileptonic events constitute the two main particularities of this physics analysis with respect to other differential measurements of top quark pair production.

\section{Analysis event selection}
\label{sec:eventselection}
The event selection consists of at least two reconstructed jets of which at least one is a b-tagged jet, satisfying requirements on transverse momentum and pseudorapidity of \ensuremath{p_{\text{T}}} > 30 GeV, $|\eta| < 2.4$. In addition, at least two charged (sub)leading leptons (electrons, muons) of opposite charge with \ensuremath{p_{\text{T}}} > 20 (25) GeV, $|\eta| < 2.4$ are required. A veto on events with additional leptons (electrons or muons) with \ensuremath{p_{\text{T}}} higher than 15 GeV is applied. Events are further separated into same-flavor and different-flavor channels. The clean selection obtained has an overall $78\%$ of signal contribution, with the largest background contributions coming from \ensuremath{t\bar{t}} other processes, single top, and Drell Yan plus jet events. 

\section{Improving \ensuremath{p_{\text{T}}^{\text{miss}}} resolution observable}
\label{sec:improvingresolutionobservable}
\begin{figure}[htp!]
    \centering
        \includegraphics[width=.49\textwidth]{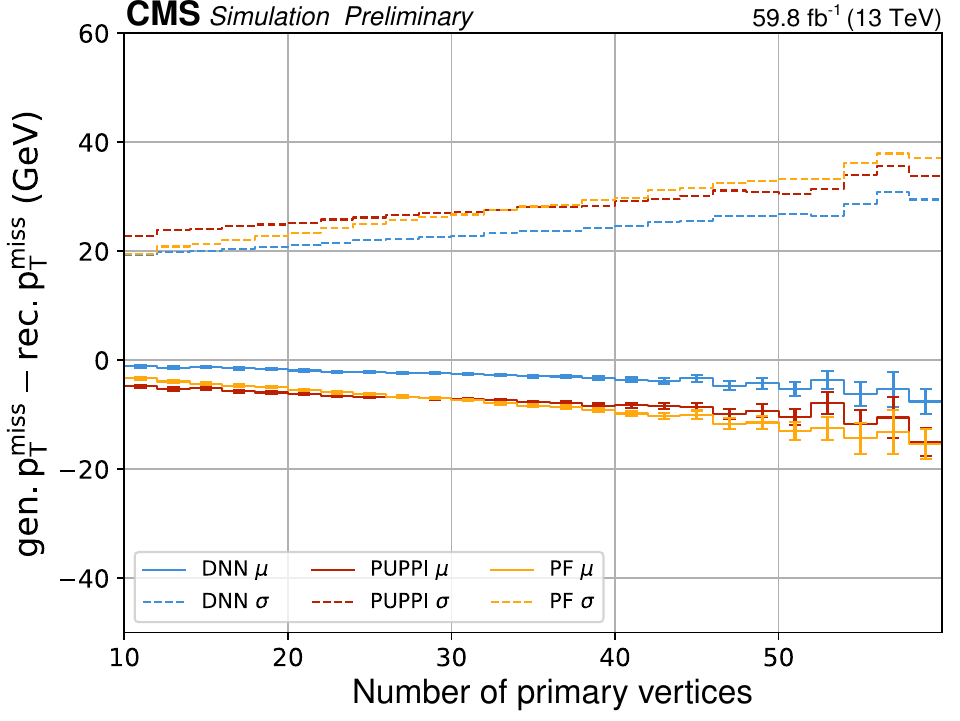}
    \caption{Difference between generated and reconstructed \ensuremath{p_{\text{T}}^\text{miss}} as a function of the number of primary vertices. The mean ($\sigma$) difference between \ensuremath{p_{\text{T,gen.}}^\text{miss}} and \ensuremath{p_{\text{T,rec.}}^\text{miss}} per bin is shown as solid (dashed) line for the \ensuremath{p_{\text{T}}^{\text{miss}}} corrected by the DNN regression (light blue), derived with PUPPI algorithm (red), and particle flow algorithm (orange).}
    \label{fig:dnnperformance}
\end{figure}

The main sources of missing transverse momentum on dileptonic \ensuremath{t\bar{t}} events come from the \ensuremath{p_{\text{T}}^{\nu\nu}} of the two prompt neutrinos produced in the dileptonic decays, non-prompt neutrinos from semileptonic meson decays in jets, and mismeasurement of particle momenta during reconstruction with the largest impact for this third source arising from mismeasurements in jets. Poor resolution or large biases of the reconstructed \ensuremath{\vec{p}_{\text{T}}^{\text{miss}}} can compromise the stability of the unfolding procedure.
A dedicated DNN regression to correct for detector effects and ensure a more accurate reconstruction of the magnitude and direction of \ensuremath{\vec{p}_{\text{T}}^{\text{miss}}} has been developed in the context of this physics analysis \cite{Abadi:2016kic, Andreassen:2019nnm}. A feed-forward, fully-connected DNN with two output nodes, mainly the x and y components of the difference between \ensuremath{\vec{p}_{\text{T,PUPPI}}^\text{miss}} and generated \ensuremath{\vec{p}_{\text{T}}^{\text{miss}}} (\ensuremath{\vec{p}_{\text{T,gen.}}^\text{miss}}) is used. PUPPI refers to the pileup-per-particle identification algorithm which allows for a reduction of the pileup dependence of \ensuremath{p_{\text{T}}^{\text{miss}}} \cite{CMS:2019ctu,Bertolini:2014bba}. The performance of the DNN regression method can be assessed in Fig.~\ref{fig:dnnperformance}. Overall, the resolution of \ensuremath{p_{\text{T}}^{\text{miss}}} corrected by the DNN regression is improved by approximately $15\%$ when compared to \ensuremath{p_{\text{T,PUPPI}}^\text{miss}}. Furthermore, the resolution of $\phi$(\ensuremath{\vec{p}_\text{T}^{\text{miss}}}) is improved by around $12\%$. As a result, the analysis profits from a finer binning in the differential measurements of the target observables given the achieved bin-to-bin migration reduction while ensuring a stable unfolding.

\vspace{-0.5cm}

\section{Systematic uncertainties}
\label{sec:sysuncertainties}
Experimental and theoretical sources of systematic uncertainties have an effect on the differential cross section measurements given their influence on the response matrix used in the unfolding procedure and the background estimation. The impact of each systematic uncertainty is evaluated individually by varying the individual uncertainty source by one standard deviation or by using alternative simulation settings. The change with respect to the nominal extracted cross section is taken as uncertainty for that particular systematic variation. Fig.~\ref{fig:systuncbreakdown} shows the breakdown of experimental and theoretical uncertainties as a function of the one-dimensional kinematic variables. The Jet Energy Scale (JES) constitutes the dominant experimental uncertainty for most bins. For the theoretical uncertainties, at large \ensuremath{p_{\text{T}}^{\nu\nu}} and lowest \ensuremath{\text{min}[\Delta\phi(\vec{p}_{\text{T}}^{\nu\nu},\vec{p}_{\text{T}}^{\ell})]} bin, the choice of tW-$\text{t}\bar{\text{t}}$ overlap removal scheme (single top DS/DR) dominates while for the highest \ensuremath{p_{\text{T}}^{\nu\nu}} and {\ensuremath{\text{min}[\Delta\phi(\vec{p}_{\text{T}}^{\nu\nu},\vec{p}_{\text{T}}^{\ell})]} there are also sizeable contributions from matrix element (ME)-parton shower (PS) matching and ME scale. 

\begin{figure}[!htb]
    \centering
        \includegraphics[width=0.7\textwidth]{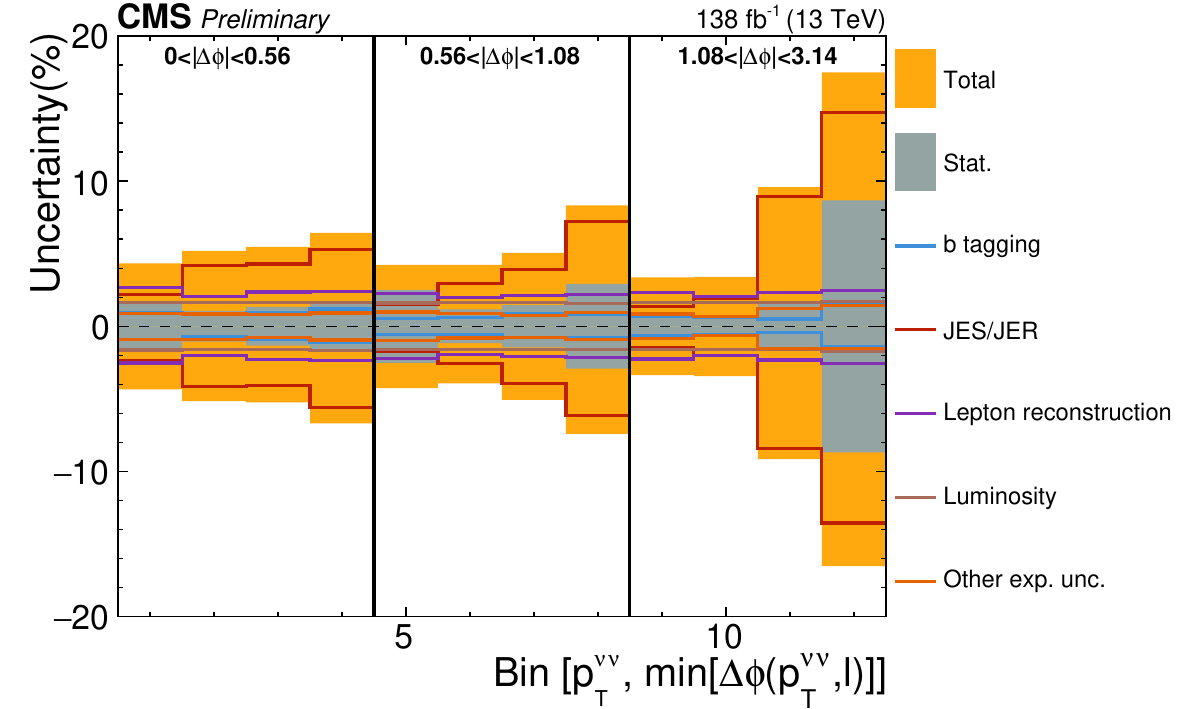} \\
        \includegraphics[width=0.7\textwidth]{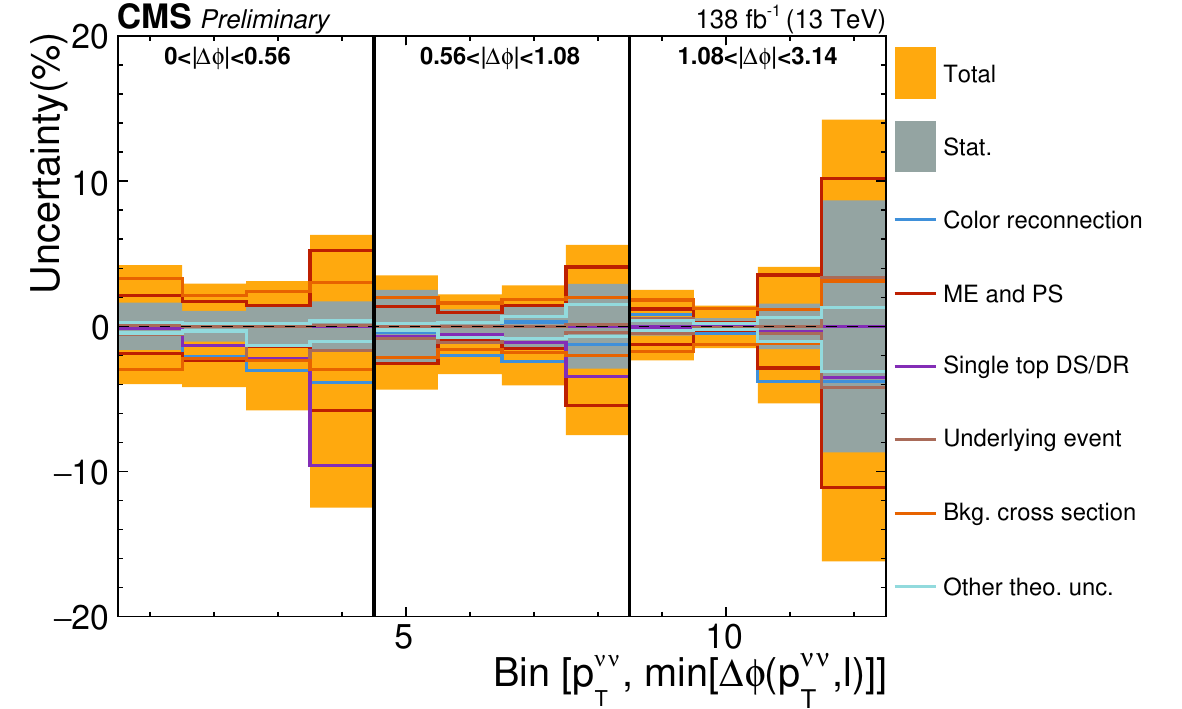}
    \caption{Breakdown of the relative uncertainties from (top) experimental and (bottom) theory uncertainties on the 2D differential cross section measurement as a function of \ensuremath{p_{\text{T}}^{\nu\nu}} and \ensuremath{\text{min}[\Delta\phi(\vec{p}_{\text{T}}^{\nu\nu},\vec{p}_{\text{T}}^{\ell})]}.}
    \label{fig:systuncbreakdown}
\end{figure}

\section{Beyond the standard model closure test}
\label{sec:bsmclosuretest}
The potential BSM contributions based on a stop pair production scenario with a stop mass of 525 GeV and a neutralino mass of 350 GeV are studied with a pseudodataset based on the nominal signal prediction plus a prediction for the stop pair production scenario. Nominal, regularized, and bin-by-bin unfolding based on the full Run 2 data-taking period is shown in Fig.~\ref{fig:closure}. The expected distributions with both neutralinos included in the particle level definition are correctly reproduced showing a reasonable sensitivity to distortions in the measured spectrum from potential BSM contributions.

\vspace{0.5cm}

\begin{figure}[htp!]
    \centering 
        \includegraphics[width=.7\textwidth]{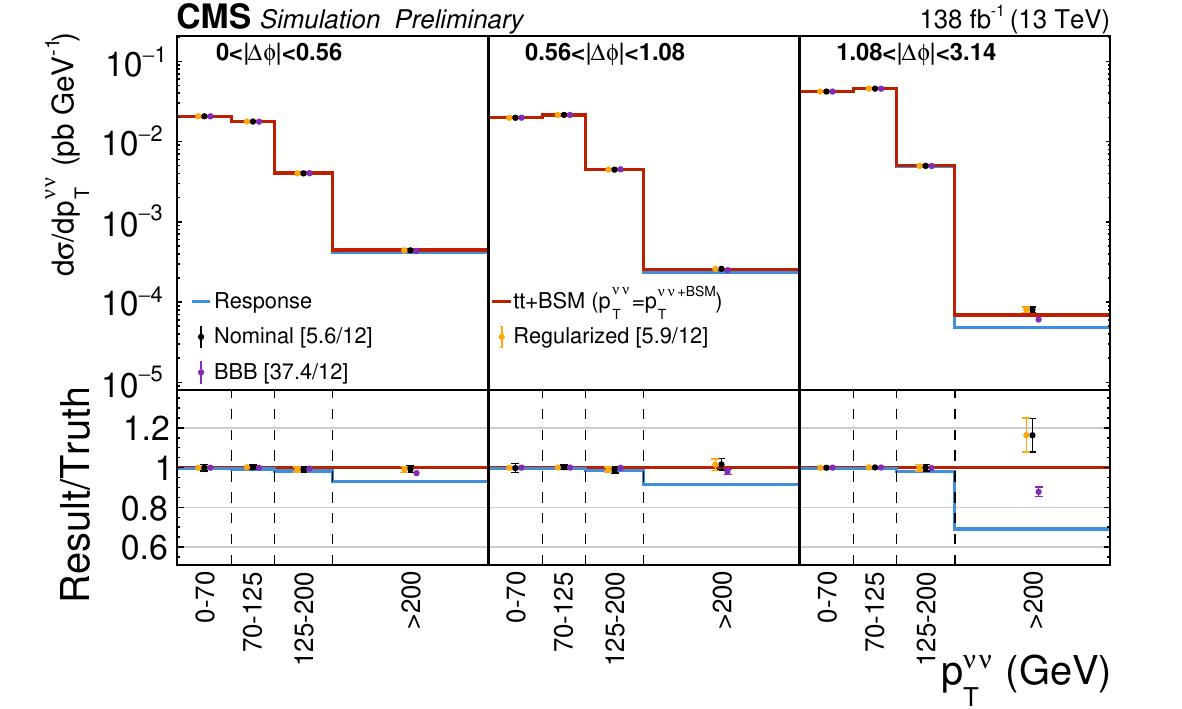}
    \caption{Result of the closure test based on simulation. The potential BSM contribution is scaled by a factor of ten. The test is performed for the 2D measurement using nominal (black), regularized (orange), and bin-by-bin unfolding (purple).}
    \label{fig:closure}
\end{figure}

\section{Differential cross section results}
The obtained differential cross sections shown in Fig.~\ref{fig:xsecs} are compared to five theoretical predictions \cite{Sjostrand:2014zea,Frixione:2007vw,Bellm:2015jjp,Czakon:2020qbd}. Small shape differences between the measured cross section and the predictions can be observed aside from the overall good agreement. In particular, the differences observed in the last bin of \ensuremath{\text{min}[\Delta\phi(\vec{p}_{\text{T}}^{\nu\nu},\vec{p}_{\text{T}}^{\ell})]} match observations from previous measurements of the azimuthal angle between two leptons, strongly correlated with \ensuremath{\text{min}[\Delta\phi(\vec{p}_{\text{T}}^{\nu\nu},\vec{p}_{\text{T}}^{\ell})]} \cite{CMS:2019nrx}.

\begin{figure}[htp!]
    \centering 
        \includegraphics[width=.7\textwidth]{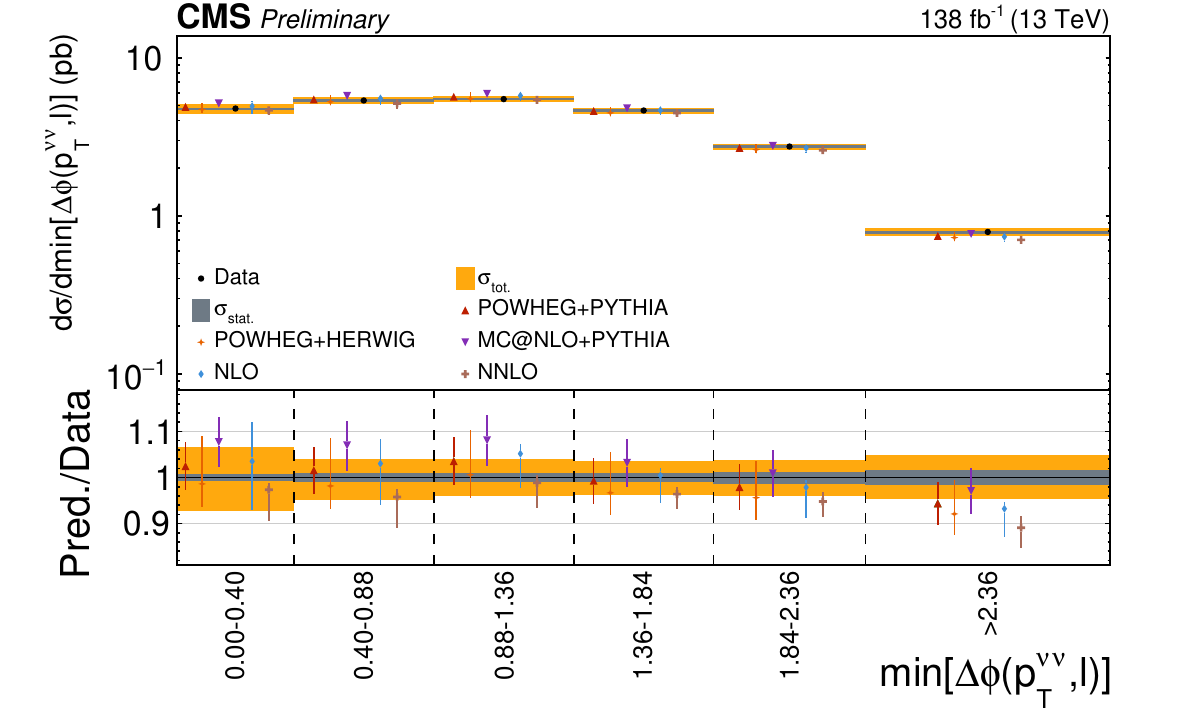}
        \includegraphics[width=.7\textwidth]{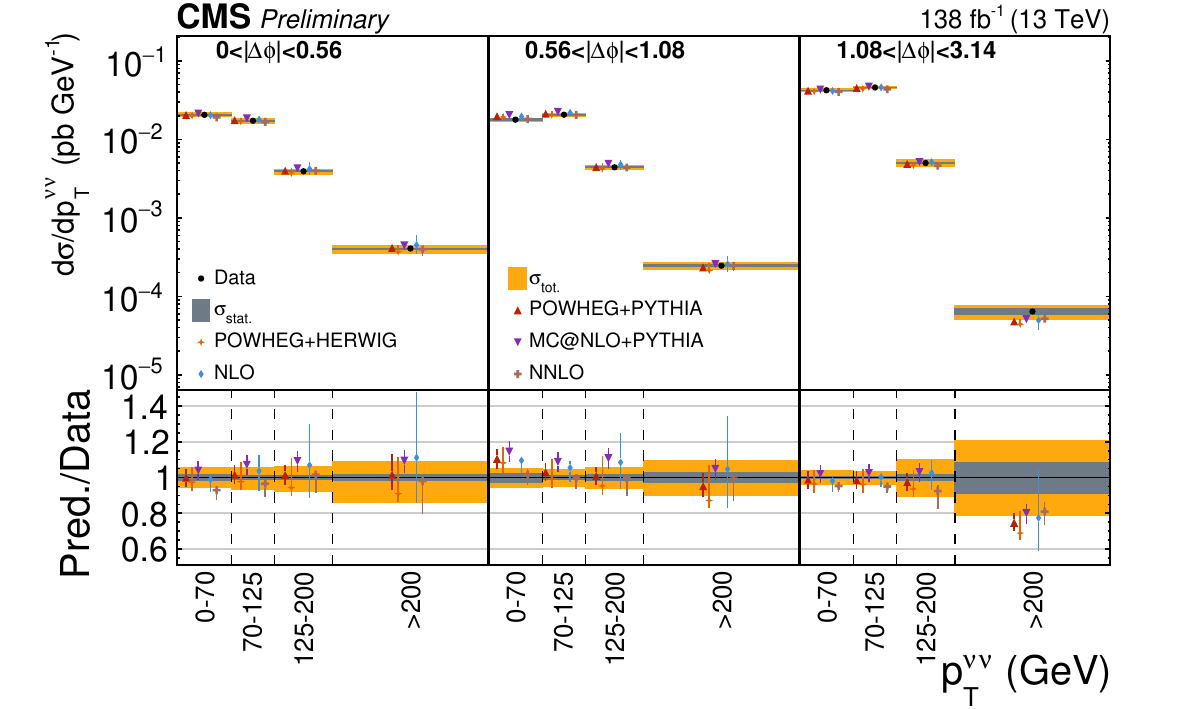}
    \caption{The measured signal cross sections (black markers) as a function of (top) \ensuremath{\text{min}[\Delta\phi(\vec{p}_{\text{T}}^{\nu\nu},\vec{p}_{\text{T}}^{\ell})]} and (bottom) both observables in two dimensions are shown. The three theoretical predictions, and the fixed-order NLO (light blue) and NNLO (brown) calculations are compared to the measurement. The total (statistical) uncertainty on the measurement is shown as an orange (dark grey) band.
    }
    \label{fig:xsecs}
\end{figure}

\vspace{-0.5cm}

\section{Summary}
\label{sec:summary}
Differential top quark pair production cross section measurements based in the dileptonic channel in proton-proton collisions have been presented, which constitute the first differential cross section measurements based on the dineutrino kinematic properties. The use of a dedicated deep neural network regression method significantly improves the missing transverse momentum resolution, with a good agreement between the different theory predictions and the measured differential cross sections observed.

\bibliography{SciPost_Proceedings_TOP2024.bib}


\end{document}